\documentclass[aps,prb,superscriptaddress,twocolumn,longbibliography]{revtex4-2}
\usepackage{graphicx,xcolor}
\usepackage{amsmath,amssymb}
\usepackage{physics}
\usepackage{dsfont}
\usepackage[colorlinks,citecolor=blue,urlcolor=blue]{hyperref}

\begin{document}

\title{
Synthetic Horizons in an Atomic Chain: Horizon-Induced Effects and Connections to Quantum Hall systems
}

\author{Ali~G.~Moghaddam}\address{Department of Applied Physics, Aalto University, P.O. Box 11000, FI-00076 Aalto, Finland}
\affiliation{Computational Physics Laboratory, Physics Unit, Faculty of Engineering and
Natural Sciences, Tampere University, FI-33014 Tampere, Finland}

\author{Viktor K\"{o}nye}
\affiliation{Institute for Theoretical Solid State Physics, IFW Dresden and W\"{u}rzburg-Dresden Cluster of Excellence ct.qmat, Helmholtzstr. 20, 01069 Dresden, Germany}
\affiliation{Institute for Theoretical Physics, University of Amsterdam, Science Park 904, 1098 XH Amsterdam, The Netherlands}
\author{Lotte Mertens}
\affiliation{Institute for Theoretical Solid State Physics, IFW Dresden and W\"{u}rzburg-Dresden Cluster of Excellence ct.qmat, Helmholtzstr. 20, 01069 Dresden, Germany}
\affiliation{Institute for Theoretical Physics, University of Amsterdam, Science Park 904, 1098 XH Amsterdam, The Netherlands}

\author{Jasper van Wezel}
\affiliation{Institute for Theoretical Physics, University of Amsterdam, Science Park 904, 1098 XH Amsterdam, The Netherlands}
\author{Jeroen van den Brink}
\affiliation{Institute for Theoretical Solid State Physics, IFW Dresden and W\"{u}rzburg-Dresden Cluster of Excellence ct.qmat, Helmholtzstr. 20, 01069 Dresden, Germany}
\affiliation{Institute for Theoretical Physics, TU Dresden, 01069 Dresden, Germany
}

\date{\today}

\begin{abstract}
We investigate the sine model, a one-dimensional tight-binding Hamiltonian with position-dependent sinusoidal hopping, and demonstrate the formation of synthetic horizons where electronic wave packets exhibit exponential slowdown. Interestingly, using the exact transformation between this model and the Harper equation, we find that analogous semiclassical horizons can emerge in a quantum Hall setup within the Harper-equation representation and at half filling for specific values of the magnetic flux. The Harper equation describes the eigenstates of a square-lattice tight-binding model subjected to a perpendicular magnetic field, as a family of one-dimensional problems parametrized by a conserved transverse quasi-momentum. We demonstrate the correspondence between the sine and Harper models by comparing their semiclassical phase-space portraits and by numerically studying wave-packet dynamics in both models. Furthermore, by applying sudden quenches to the sine model's hopping profile, we observe the emergence of states with an effective thermal Fermi–Dirac distribution characterized by a corresponding Unruh temperature. Our numerical calculations reveal a non-universal behavior of this temperature, suggesting the involvement of mechanisms beyond a simple low-energy description. 
\end{abstract}

\maketitle

\section{Introduction}

One of the fascinating aspects of quantum condensed matter systems lies in the remarkable opportunity to tailor and manipulate them 
such that they correspond to exotic theoretical models in extreme limits. This allows us to mimic seemingly unrelated physical phenomena within these systems. A prime example is the Sachdev-Ye-Kitaev (SYK) model, initially conceived as a quantum spin system with random, all-to-all interactions, which was later found to have surprisingly profound connections to quantum gravity \cite{Sachdev1993,kitaev2015simple,kitaev2018,Maldacena2016prd,Maldacena2016}. Another more tangible example is the emergence of Lorentz symmetry in the low-energy description of certain materials, which gives rise to relativistic physics in their continuum limit. Notably, three-dimensional Dirac and Weyl semimetals, as well as two-dimensional (2D) graphene, are accurately described by the Dirac equation in their low-energy regimes \cite{Armitage2018}. Importantly, Lorentz invariance can also hold in systems with a massive Dirac spectrum, where a gap opens but the relativistic structure is preserved \cite{Seradjeh2022}.

Beyond the quantum field theoretic aspects of the above systems, which exhibit features such as the chiral anomaly and the Klein paradox, they offer the exciting possibility of simulating the quantum dynamics of Dirac fermions in curved spacetimes \cite{Volovik:2016kid,volovik2003universe,Weststrom2017,hartnoll2011horizons,guan2017artificial,huang2018black,Ojanen2019,Meng2023scipost,Konye2023lensing,Konye2022weyl}. This can be achieved by carefully engineering the material properties to possess an effective position-dependent description \cite{kedem2020black,Morice2021,morice2022quantum}. Consequently, it becomes feasible to synthesize analogs of general relativistic objects, such as black holes, providing an accessible experimental platform to investigate specific aspects of black hole physics. Numerous setups have been proposed and even implemented to explore this so-called analog gravity within condensed matter systems, including superfluids \cite{hu2019Unruh,steinhauer2016,Nissinen2020,Barcelo2001,barcelo2011analogue,Carusotto2008,nguyen2015}, and fermionic or spin systems \cite{Calabrese2017,Minar2015,Jafari2019,Staalhammar_2023,Horner2023,daniel2025optimally}, optical and magnonic systems \cite{philbin2008,duine2017}, and electronic circuits \cite{kollar2019hyperbolic,Boettcher2020,Boettcher2022}.

In this paper, we introduce the sine model, a one-dimensional (1D) tight-binding Hamiltonian characterized by spatially varying hopping amplitudes following a sinusoidal pattern, $\sin(\pi\alpha n)$. This model exhibits horizon physics at the spatial locations where the hopping strength vanishes. In this context, $n$ corresponds to the lattice site index, and the parameter $\alpha = P/N$ takes rational values, where $N$ is the system size and $P$ dictates the number of periods in the sinusoidal modulation. In the low-energy regime, the sine model is equivalent to a 1D chain with a linearly increasing hopping strength~\cite{Morice2021,morice2022quantum}, which has been shown to allow a direct condensed matter analog for the instantaneous creation of horizons in a gravitational setting~\cite{Mertens2022}. 

While the sine model holds significant intrinsic interest, as evidenced by its potential realization in experimental platforms such as ultracold atoms in optical lattices with modulated inter-site coupling and in metamaterials, it also exhibits a profound relationship with the integer quantum Hall effect in a square lattice through its description in terms of the Harper's equation. Indeed, a specific gauge transformation establishes a mapping between the Harper model~\cite{Wiegmann1994,krasovsky1999bethe,Marra2025} and the sine model, and vice versa. In the context of the quantum Hall effect, the magnetic flux $\Phi$ per unit cell, when expressed in units of the flux quantum $\Phi_0$ as $\alpha = \Phi/\Phi_0$, determines the key parameter $\alpha$. In typical scenarios, this corresponds to a regime of weak magnetic field and large magnetic length. Furthermore, at the electron-hole symmetric point of half-filling, this setup leads to the closure of the energy gap and the emergence of zero-energy modes~\cite{Wen1989}, resulting in a non-quantized Hall conductance~\cite{Hatsugai1990}. 

It is important to note that the Harper model provides an effectively 1D description for each value of the transverse quasi-momentum. Consequently, while synthetic horizons can be studied within this framework, their manifestation in the full 2D quantum Hall system retains this effectively 1D character owing to translational symmetry along one direction and given that the considered representation also respect this symmetry. It should be noted that, while both the explicit form of the wavefunctions and the decomposition into a family of effectively 1D channels (leading to Harper’s equation) rely on a specific gauge choice, the underlying physical effects are gauge invariant in the full two-dimensional system.
Now, in direct correspondence with the sine model, we also demonstrate the formation of semiclassical horizons in the Harper model, where electron wave packets exhibit exponential slowing down on the separatrices of its phase portrait. Here, the term ``semiclassical horizon” refers to the description of horizons in the continuum limit of the sine and Harper models, where the lattice system is approximated by continuous variables and classical dynamics.
Numerical simulations of wave packet propagation in the quantum model corroborate these analytical findings and their limitations.

Furthermore, we demonstrate how sudden quenches of the hopping profile in the sine model, between configurations with different periodicities, can lead to the emergence of thermal-like states. Starting from the ground state of the pre-quench Hamiltonian, the system immediately after the quench appears as a thermal Fermi--Dirac distribution when viewed from the perspective of the post-quench Hamiltonian. This form of thermalization can be interpreted in the context of Unruh physics and its associated temperature, since the post-quench Hamiltonian, particularly in its low-energy sector, can be regarded as the initial Hamiltonian seen from the perspective of an accelerated observer.
Specifically, starting from the quantum ground state of the sine model at zero temperature for a given $P$, and by abruptly increasing the number of periods, or equivalently the number of horizons, across the chain by an integer $\Delta P$, we observe this thermalization. This process, driven by the formation of new horizons, leads to an Unruh temperature given by $k_B T /E_F = c \Delta P$, which originates from the entanglement present in the pre-quench multi-particle ground state across the newly formed horizon. Our direct numerical calculations yield the Unruh temperature and the constant of proportionality $c$, revealing a non-universal behavior. This non-universality indicates that the flux quench in the sine model involves processes that extend beyond the description provided by the effective low-energy Hamiltonian of a chain with linearly increasing hopping.

\section{Horizon physics in the Harper and sine Models}

\subsection{Sine model}

We begin by introducing the \emph{sine hopping model}, which is defined on a 1D lattice with sites $n \in [0, N-1]$ and described by the Hamiltonian  
\begin{align}
    \hat{H}_S = \sum_{n=0}^{N-2}  2\sin(\pi \alpha n) \hat{c}^\dagger_n \hat{c}_{n+1} + h.c. ,
    \label{eq:sine}
\end{align}
where $\hat{c}^\dagger_n$ and $\hat{c}_{n}$ are the fermionic creation and annihilation operators at site $n$, respectively. The parameter $\alpha$ is a rational number, which can be expressed as $\alpha = P/N$, where $P$ is the number of periods in the sine modulation and $N$ is the system size. The Hamiltonian \eqref{eq:sine} describes a tight-binding model with nearest-neighbor hopping, where the hopping amplitude varies sinusoidally with the lattice site index $n$. The eigenvalue equation for the sine model is thus given by set of coupled equations
\begin{align}
   2\sin\left( \pi \alpha n \right) \phi_{n-1}
   + 2\sin\left( \pi \alpha (n+1)\right) \phi_{n+1} = \epsilon\phi_n,
\end{align}
for the eigenvalue $\epsilon$.

In the vicinity of points where the hopping amplitude vanishes, the sine model exhibits a synthetic horizon similar to that in linearly increasing hopping models ~\cite{Morice2021,morice2022quantum,Mertens2022}.
To see this, it is helpful to first consider the sine model in the continuum limit (see Appendix \ref{sec1}), where the Hamiltonian can be expressed as
\begin{equation}
    \hat{\mathcal{H}}_{S} = 4 \cos(\hat{p}) \sin(\pi \alpha \hat{x}).
\end{equation}
Hence, the hopping amplitude vanishes at points
\begin{equation}
    x_h \equiv \frac{m}{\alpha} ,
\end{equation}
where $m \in \mathbb{Z}$, and $x_h$ is a continuous parameter that coincides with $n_h$ at integer values. 
In the vicinity of such points, the sine model can can be approximated by a tight-binding chain with linearly increasing hopping strengths:
\begin{equation}
    \epsilon \phi_n=\pi \alpha \:
    \left[
    n\phi_{n-1} + (n+ 1)\phi_{n+1} 
    \right],
\end{equation}
which harbors the synthetic horizons.

\subsection{Harper's model}

Next, we consider the Landau-gauge Schr\"odinger equation for 2D lattice electrons in a magnetic field which reduces, after a Fourier transform of one of the spatial directions, to the well-known 1D Harper's equation~\cite{harper1955single,hofstadter1976energy, tong2016lectures,Wiegmann1994,krasovsky1999bethe} 
\begin{align}
    \psi_{j-1} + 2 \cos(2 \pi \alpha j + \theta) \psi_j + \psi_{j+1} = \epsilon \psi_j.
    \label{eq:Harper}
\end{align}
Here, the wave function $\psi_j$ is defined on discrete lattice sites $j \in [0,N-1]$, the energy is denoted by $\epsilon$, and $\theta \in \mathbb{R}$ is the quasi-momentum corresponding to the Fourier transformed spatial direction. Finally, $\alpha$ is equal to the magnetic flux through a lattice cell divided by the magnetic flux quantum. Note that the energy $\epsilon$ is measured in units of the hopping amplitude $\lambda$, so that a half-filled band with particle-hole symmetry has Fermi energy $E_F=2 \lambda$.

\begin{figure}
    \centering
    \includegraphics[width=8.6cm]{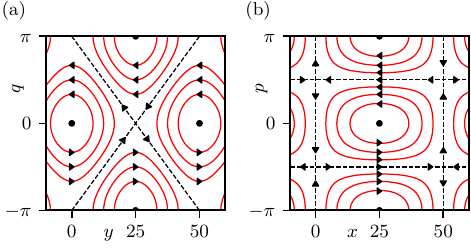}
   \caption{\label{fig:SCminussin}Phase-portraits for Harper's equation (a) and the sine model (b), with $\alpha=1/50$ and $\theta=0$. The center for the circular solutions is given by the black dot at $q=0$ and $p=0$. The portraits are related by a rotation of $\pi/4$ and a shift of the centers, as well as a rescaling. The black dashed lines indicate the separatrices.}
    \label{fig:enter-label}
\end{figure}

The continuum form of the Harper equation is given by
\begin{align}
    \hat{\mathcal{H}}_{H} & = 2\cos(\hat{q}) + 2\cos(2\pi \alpha \hat{y} + \theta),
\end{align}
where $\hat{q}$ and $\hat{y}$ denote the momentum and position operators, respectively (see Appendix~\ref{sec1}). As mentioned in the introduction and detailed later, an exact mapping exists between the sine model and the Harper equation. In the continuum limit, this mapping can be readily established through the following canonical transformation:
\begin{align}
    \hat{q} &= -\hat{p} + \pi \hat{x} \alpha - \frac{\pi}{2}, \\
    2\pi \alpha \hat{y} &= \hat{p} + \pi \hat{x} \alpha - \theta - \frac{\pi}{2}.
\end{align}
The connection between these two models can be elegantly visualized through the phase portraits of their respective continuum Hamiltonians in the classical limit. The phase portraits for the Harper and sine models are depicted in Fig.~\ref{fig:SCminussin}, each comprising two lattices of closed orbits separated by separatrix lines (dashed in black). Consistent with the canonical transformation provided above, the two phase portraits are related by a combination of shift, rescaling, and rotation.

Importantly, the canonical transformation between the continuum-limit Hamiltonians preserves the classical trajectories, which exhibit an exponential slowdown characteristic of horizon physics in both the sine and Harper models. Restricting to a separatrix, the Heisenberg equations of motion yield the following time dependence for the classical expectation value of position (as detailed in Appendix~\ref{sec1}):
\begin{align}
    y(t) = \frac{1}{\pi \alpha} \left[ \arctan\Big(e^{-4\pi\alpha(t-t_0)}\Big) - \theta/2 \right]+\frac{1}{2\alpha}.
    \label{eq:HarperSlowDown}
\end{align}
As anticipated for a separatrix in a classical phase portrait, the position exponentially slows down as it approaches the intersection with another separatrix.

\subsection{Exact map Between the Harper and sine models}

As previously mentioned, an exact mapping exists between the Harper equation and the sine model~\cite{Wiegmann1994,krasovsky1999bethe}. This mapping holds for the discrete lattice versions of both models (i.e., beyond the continuum limit) under the following conditions, detailed in Appendix~\ref{sec2}:
\begin{enumerate}
    \item $P$ and $N$ are coprime.
    \item $\theta = (2m'+1)\frac{\pi}{N}$ with $m'\in\mathbb{Z}$.
\end{enumerate}
This mapping is crucial for connecting the numerical simulations of wave packet dynamics in the Harper model to the horizon physics observed in that model, and similarly for the sine model. Subject to the conditions above, an eigenfunction $\phi_m$ of the sine model is mapped to an eigenfunction $\psi_n$ of the Harper equation by:

\begin{equation}
\begin{split}
    \psi_n = &\frac{1}{N}\sum_{k}^{N-1} 
     e^{ik(\theta - \pi)}
     \:    e^{
     i\pi\alpha\left[2nk+ (k+\frac{1}{2})^2\right]}
    \\    &
    \qquad\quad\sum_{m}^{N-1} 
    i^{-m}\: e^{i
    \pi\alpha\left[ 2km + \frac{m(m+1)}{2}\right]}\:
 \phi_m.
\end{split}
\end{equation}

It should be noted that the boundary conditions play an important role in establishing this correspondence. As detailed in Appendix~\ref{sec2}, the choice of boundary conditions for the sine model is largely inconsequential due to the vanishing hopping amplitudes at the chain edges, which effectively decouples the boundaries from the bulk. In contrast, for the Harper model, we assume periodic boundary conditions throughout, which are essential for ensuring the validity of the exact mapping between the two models.

\section{Wave packet dynamics near the horizon}
It has been established previously \cite{morice2022quantum, Morice2021, Mertens2022,Konye2023lensing,Konye2022weyl} that models with position-dependent nearest-neighbor hopping can have dynamics of zero-energy wave packets that coincide with the dynamics of Dirac fields in a static curved spacetime. The semiclassical mapping also manifests in the dynamics of coherent wave packets in the current model, which we will explore by numerical simulation.

While the original quantum Hall system is inherently 2D, the Harper model can be effectively reduced and treated as a 1D model through an appropriate gauge choice. In this gauge, one spatial direction is taken to be periodic and is Fourier transformed, yielding a conserved quasi-momentum parameter $\theta$, while the other direction remains open. Harper’s equation then describes the dynamics along this open direction, with $\theta$ appearing as a fixed parameter. 
This formulation shows that the Harper model represents the quantum Hall system as a family of effectively 1D problems, each labeled by a fixed value of the transverse quasi-momentum $\theta$. 
Therefore, in our simulations, we focus on the evolution of wave packets at fixed $\theta$, since it remains constant under the dynamics. 
Consequently, although the underlying system is genuinely 2D, the relevant wave-packet dynamics take place entirely along a single spatial direction for each $\theta$.
As a result, the time evolution occurs purely along the remaining spatial direction and maps directly onto the 1D sine model introduced earlier. From this perspective, the discussion of synthetic horizons within the Harper framework naturally captures horizon physics in the quantum Hall system, albeit in an effectively 1D setting determined by the choice of gauge and the conserved quasi-momentum.

It is important to emphasize that, although the decomposition of the 2D quantum Hall problem into independent 1D channels and the explicit form of the wavefunctions depend on the choice of gauge, the underlying physical effects do not. In fact, one can note that while the form of the Hamiltonian changes under a gauge transformation, this corresponds only to a unitary transformation and therefore does not affect the probability density of wave packets. In particular, the horizon-like behavior of wave packets, such as their slowing down and turning points, is thus gauge invariant and persists in the full two-dimensional system. This is somehow similar to the usual treatment of edge states, where a convenient gauge is chosen to simplify the description, while physical observables such as conductivity remain gauge invariant.

\begin{figure}
\includegraphics[width=\linewidth]{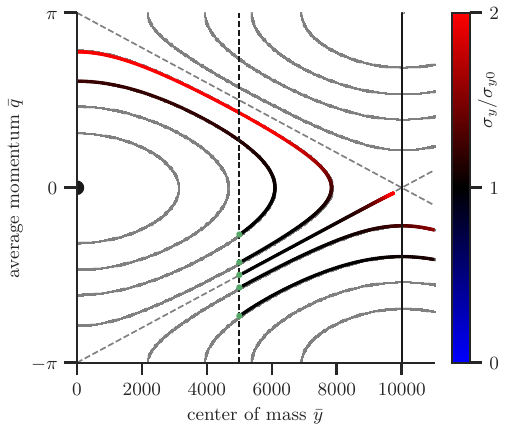}
\caption{ \label{fig:PhaseSpace} Time evolution of wave packets in Harper's model, Eq.~\eqref{eq:Harper}, where the initial wave packet has a momentum between $-\pi$ and $\pi$. All wave packets start at the dashed black line at lattice point $5000$. The color of the lines denotes the ratio of the standard deviation of the position of the wave packet $\sigma_y$ and its value at the beginning of the simulation $\sigma_{y0}$. 
The vertical black line marks the position of the horizon encountered along the separatrix, which lies at $y_h = 10000$ with $P=1$ and $N=20 000$, while the gray lines mark semiclassical trajectories.
}   
\end{figure}

For Harper's model the horizon encountered on the separatrices is located at $y_h \equiv \frac{2m+1}{2\alpha} - \frac{\theta}{2\pi \alpha}$ for $m\in\mathbb{Z}$, corresponding to the crossing point of the dashed black lines in the semiclassical phase portrait of Fig.~\ref{fig:SCminussin}(a). For a coherent wave packet around a given initial momentum and position, the phase portrait for the evolution of its center of mass is shown in Fig.~\ref{fig:PhaseSpace}.  
This figure shows that the quantum evolution of the wave packet's center of mass matches the previous semiclassical analysis. At initial momenta close to $q=0$, harmonic oscillations are recovered, while for momenta on the separatrices, the evolution asymptotically approaches the horizon at $q=0$ and $y= y_h$ while exponentially slowing down, until the wave packets become delocalized as finite size effects take over. We also note that these wave-packet dynamics and horizon effects can be directly related to appropriately prepared wave packets in a 2D quantum Hall system, provided they have a well-defined transverse quasi-momentum.

\begin{figure}[t]
\includegraphics[width=\linewidth]{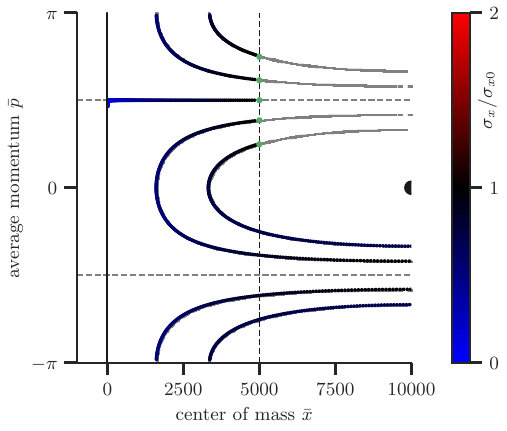}
\caption{ \label{fig:PhaseSpaceSine} Time evolution of wave packets in the sine model, Eq.~\eqref{eq:sine}. The plot style and parameters are identical to those used in Fig.~\ref{fig:PhaseSpace}. In this model, the wave packets exhibit clear localization upon approaching the horizon, as indicated by a reduction in the standard deviation $\sigma_x$, which characterizes the average spatial width of the wave packet. The parameters used here are the same as those in the corresponding simulations for the Harper model.
} 
\end{figure}

For both Harper's equation and the sine model, a set of synthetic horizons arise, as witnessed by wave packets exponentially slowing down and never reaching $x_h$ or $y_h$ (for a detailed calculation, see Appendix \ref{sec1}). 
The main difference between the two models is visible when we consider the evolution of $q$ as the wave packet approaches the horizon. For the sine model, starting from $p = \pi/2$, the momentum stays constant over the whole trajectory. But for Harper's equation, $d\hat{q}/dt$ only becomes zero at the horizon: it changes along the trajectory. This is due to the canonical transformations which render horizontal lines in the phase portrait of one model diagonal in that of the other.

For comparison, the corresponding wave packet dynamics for the sine model are shown in Fig.~\ref{fig:PhaseSpaceSine}. The wave packets initially localize near the horizon, consistent with earlier findings in Ref.~\cite{Morice2021}. 
This contrasting behavior in the two models can be attributed to the fact that the roles of position and momentum get mixed while mapping between the two models. While the sine model clearly exhibits purely spatial localization of wave packets near the horizon, the mixing of position and momentum in the mapping to the Harper model implies a spatial spread of the packets approaching the horizon in that model.

Returning to the sine model, although the wave packet centers initially follow classical trajectories along the separatrix, due to the quantum effects at late times, 
the wave packet cannot continue slowing down and becoming indefinitely localized.
As the wave packet's spatial width shrinks and becomes comparable to the lattice spacing, the semiclassical approximation ceases to be valid. Beyond this point, unitary quantum evolution leads to the disintegration of the wave packet near the horizon, a behavior driven by the high-energy cutoff inherent to the lattice structure~\cite{Morice2021}.

\section{Thermalization in the sine model}

We have shown the correspondence between the sine model and Harper's equation in terms of their synthetic horizons, both semiclassically and through wave-packet dynamics. This sets the stage to consider the emergence of effective thermal-like behavior in the quantum state of free fermions following a sudden change in the periodicity of the sine model, in analogy to the behavior previously observed after a quench in the linear hopping model~\cite{Mertens2022}.

Thermal radiation in gravitational settings is well known in the context of Hawking and Unruh effects, where the presence of a horizon renders part of the ground state inaccessible to observers on each side of the horizon~\cite{hawking1974nature,Unruh1976}. This leads the quantum vacuum to appear as a mixed thermal state to such observers~\cite{Israil1976,Bombelli1986,takagi_1986,Frolov1998,mann2015}. While the sine model is not fundamentally relativistic and does not involve real acceleration or observers, it exhibits emergent relativistic behavior in its low-energy sector. In this context, a sudden quench between two Hamiltonians can be interpreted, at the level of an effective field-theoretic description, as relating two frames analogous to inertial and accelerated perspectives. Hence, as an alternative route to understanding the emergence of thermal quantum states associated with horizons, we therefore study this horizon-induced, quench-generated form of effective thermalization in the sine model and show that it reproduces key features of the thermal behavior encountered in gravitational settings.

To demonstrate the emergence of this effective thermal-like behavior, we consider an instantaneous increase in the number of periods $P$ that creates one or more synthetic horizons in the lattice model with sinusoidal hopping. 
The effect of this change is characterized by calculating the number of excitations present in the system before and after the horizon's formation, in terms of instantaneous eigenstates of the Hamiltonian. Such a boost results in an emergent temperature in the sine model equal to the analog Unruh temperature. 
This has been previously shown for an $N$-site chain with homogeneous hopping quenched to one with linearly increasing hopping, which gives rise to the Unruh temperature $T = \frac{2 \lambda}{N \pi}$ for large $N$ (considering $k_B=1$),
with $\lambda$ indicating the slope of the linear dependence of the hopping parameter 
\cite{Mertens2022}. The  underlying reason for the excited state distribution to be precisely thermal was shown in that case to be that the post-quench Hamiltonian matches the entanglement Hamiltonian of the pre-quench system.
The connection between the post-quench Hamiltonian and entanglement Hamiltonian corresponding to the ground state of the pre-quench system also applies to the sine model in at least one type of quench: it has been shown that the half-system entanglement Hamiltonian of a finite ring of size $L$ and under periodic boundary conditions approaches a position-dependent Hamiltonian density $\sin(2\pi x/L){\cal H}$ where ${\cal H}$ denotes the original system Hamiltonian density \cite{Klich2013entanglement,Cardy2016,Eisler2018,Eisler2019}. These works, which are based on $(1+1)D$ conformal field theory and numerical evaluations of tight-binding chains, provide extensions of the Bisognano-Wichmann theorem relating the entanglement Hamiltonian of a generic relativistic quantum field to the boost operator \cite{BW1975,Dalmonte2022,Moghaddam2022}.

\begin{figure}
    \centering
    \includegraphics[width=\linewidth]{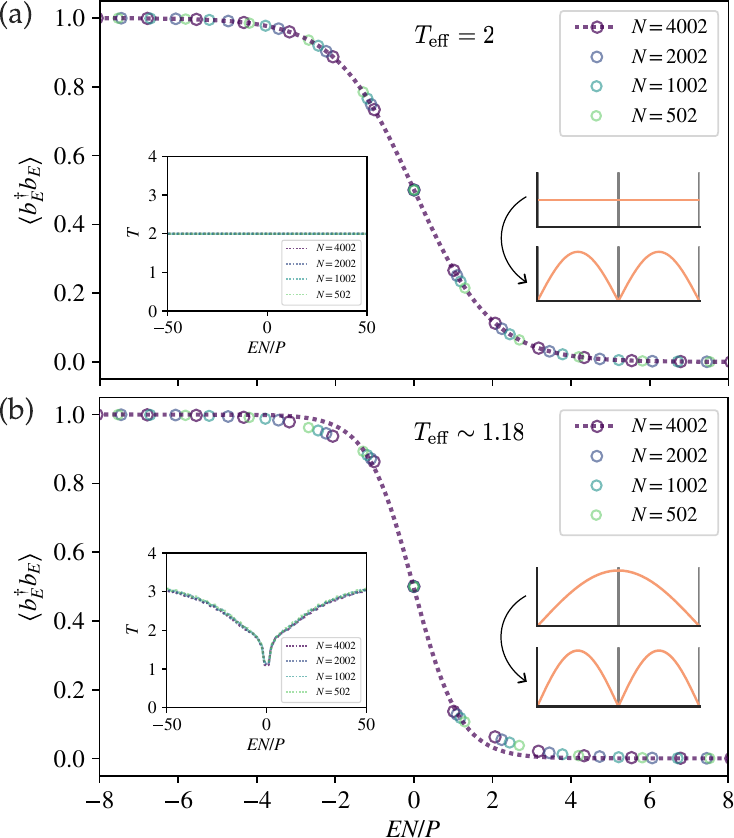}
    \caption{   
    Expectation values of the occupation number $\langle b_E^\dagger b^{\phantom\dagger}_E\rangle$ for post-quench modes, together with corresponding fits using a thermal FD distribution.
    Panel (a) corresponds to a quench from the constant hopping model to the sinusoidal hopping model with $P=2$. Panel (b) shows a similar plot but for a quench between sinusoidal hopping models from $P=1$ to $P=2$. The insets show the variation of the energy-dependent effective temperature $T(E)$, assuming a generalized thermal form $[1+\exp(E/T(E))]^{-1}$.
    While panel (a) displays an exact thermal form, panel (b) exhibits deviations from the exact FD form, evidenced by the non-constant effective temperature $T(E)$ in the inset of (b).
    }
    \label{fig:QP1}
\end{figure}

Considering a particular quench from a 
uniform hopping model 
at half-filling and under periodic boundary conditions to a sine model with $P=2$ with a pair of horizons, an exactly thermal distribution is obtained whose Unruh temperature is $T = 2\lambda/ N = E_F \Phi/\Phi_0$.
This has been illustrated in Fig. \ref{fig:QP1}(a) where the post-quench excitation spectrum is shown for different lengths.
In this case, there is an exact mapping to the entanglement Hamiltonian as mentioned earlier, which results in an 
exact Fermi--Dirac (FD) distribution for the expectation value of the  post-quench modes' occupation number $\langle b^\dag_E b^{\phantom \dagger}_E \rangle$. Here, $b^\dag_E$ and $b^{\phantom \dagger}_E$ correspond to the creation and annihilation operators of a post-quench eigenstate with energy $E$ (for details on how this is calculated see Appendix \ref{sec3}).

\begin{figure}[htp]
    \centering 
    \includegraphics[width=0.9\linewidth]{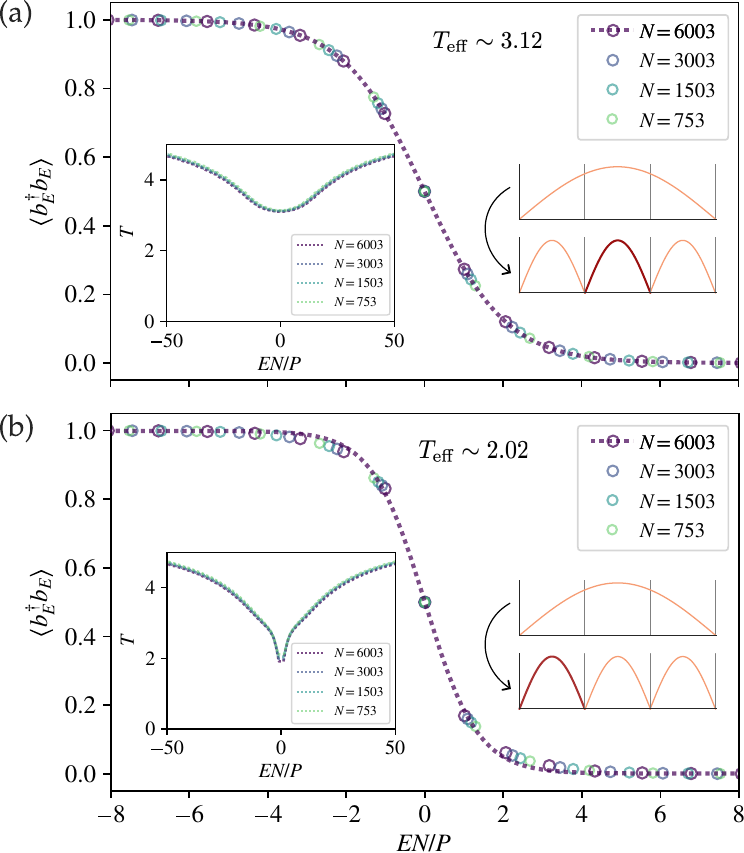}
    \caption{Expectation values of the occupation number $\langle b_\omega^\dagger b_\omega^{\phantom\dagger} \rangle$ for post-quench modes are shown, similar to the results in Fig.~\ref{fig:QP1}, but for a quench between sinusoidal hopping models from $P=1$ to $P=3$. Panels (a) and~(b) display the results for the average mode occupations in the middle and side regions, respectively, as illustrated in the two schematic insets.
    We observe that the middle region has a higher effective temperature, and its distribution is closer to a FD form compared to that of the side regions.
    \label{fig:QP2}}
\end{figure}

We note that the lattice sizes used in Fig.~\ref{fig:QP1} are chosen to be multiples of two but not four, such that each causally connected post-quench region contains an odd number of sites, ensuring the existence of a post-quench zero-energy mode.
As discussed in detail in Appendix~\ref{sec3}, for lattice sizes that are multiples of four, the thermal behavior is less accurately reproduced for finite sizes, but it asymptotically approaches the thermal form as the lattice length increases.
This type of even-odd effect, which depends on the number of sites (and thus states) within each post-quench sub-region, is a generic feature that persists even for large but finite system sizes.

Going beyond the quench from flat space to $P=1$, we also numerically examine the thermalization following a sudden quench from $P=1$ to $P=2$ as shown in Fig.~\ref{fig:QP1}(b). The resulting distribution in this case is approximately thermal, as indicated by slight deviations from the Fermi--Dirac distribution. These deviations are also evident in the energy-dependent temperature $T(E)$, derived by applying the inverse Fermi--Dirac function to the numerically obtained distribution. The effective temperature obtained from fitting is approximately half of that observed in Fig.~\ref{fig:QP1}(a).
The deviation from pure thermalization is expected because the entanglement Hamiltonian of the pre-quench state, characterized by a sine profile, no longer perfectly matches the post-quench Hamiltonian. Nevertheless, the Fermi--Dirac function remains a very good approximation, with a very small fitting error of $\chi \sim 0.05 \ll 1$ for $N=4002$, which decreases with increasing $N$ \cite{footnote-2}.

\begin{figure*}[t!]
    \centering
    \includegraphics[width=0.9\linewidth]{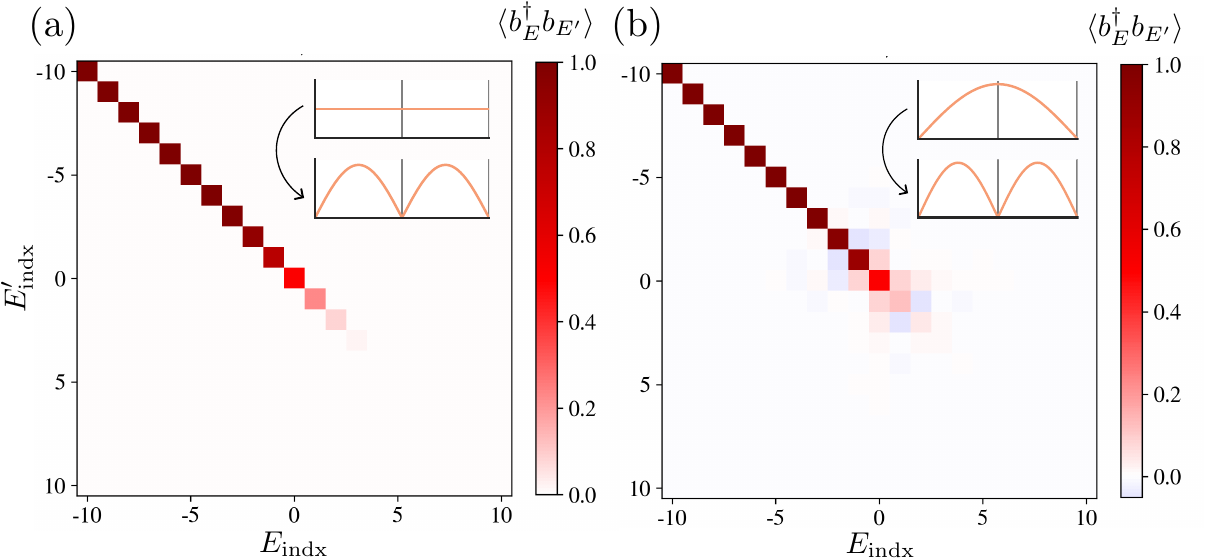}
    \caption{
    Correlation matrix elements for post-quench modes, $\langle b_{E}^\dagger b_{E'} \rangle$. The comparison is made between (a) a quench from $P=0$ to $P=2$, and (b) a quench from $P=1$ to $P=2$, as illustrated in the insets. In panel (a), we observe a perfect thermal distribution with no off-diagonal correlations. In contrast, panel (b) shows some nonzero off-diagonal correlations around zero energy, along with deviations from the perfect thermal form for the diagonal elements, as discussed in the main text. The results are obtained for a lattice size of $N=1002$. Note that $E_{\rm indx}$ and $E'_{\rm indx}$ represent the indices of the eigenvalues of the post-quench Hamiltonian for a sub-region, with $E_{\rm indx}=0$ corresponding to the zero-energy state.
    }
    \label{fig:correlations}
\end{figure*}

Additionally, we note that while the energy-dependent temperature $T(E)$ exhibits significant variation across the spectrum, between low- and high-energy states, the effective temperature obtained from a single-parameter fit aligns most closely with $T(E)$ in the low-energy regime. 
Since $T(E)$ is not obtained via fitting, but rather derived directly from the energy-resolved Fermi--Dirac distribution, this indicates that the approximate thermal behavior following the quench is primarily governed by the low-energy modes. At higher energies, deviations from a perfect thermal distribution become more pronounced. 

This energy dependence highlights the non-universal form of the thermal behavior we observe. While the low-energy modes follow a thermal distribution that can be understood from the linear approximation near the horizon, true thermalization only happens in certain quench protocols, such as when going from a uniform to a sine-modulated hopping profile. In these cases, the post-quench Hamiltonian closely matches the entanglement Hamiltonian of the pre-quench ground state near the horizon. However, at higher energies, the full details of the Hamiltonians become important, leading to deviations from perfect thermalization. These high-energy effects explain why the thermalization behavior is not universal and why the effective temperature depends on energy in more general situations.

In the following, we consider a larger boost, for example, increasing from \( P = 1 \) to \( P = 3 \), under which the post-quench system develops multiple horizons and distinct causally disconnected subregions. By analyzing the mode occupation and effective temperature within each region bounded by adjacent horizons, we find that each subregion acquires its own characteristic temperature. Additionally, we show that deviations from a perfectly thermal distribution also manifest as small but non-vanishing off-diagonal correlations, $\langle b_E^\dagger b^{\phantom\dagger}_{E'} \rangle$, near $E, E' \approx 0$.

\subsection{Regions with different temperatures} 
An interesting situation arises by considering, for instance, boost from $P=1$ to $P=3$ where the pre-quench model contains qualitatively different regions, as shown in Fig.~\ref{fig:QP2}(a). As the post-quench model contains multiple horizons, we separately consider the distribution of the mode occupations between each pair of neighboring horizons. Each of the disconnected post-quench regions is characterized by an individual temperature.
In the pre-quench model, the three post-quench regions marked in Fig.~\ref{fig:QP2}(a) are inequivalent. The middle region spans the approximately flat part of the $P=1$ sinusoidal hopping profile, while the outer regions contain approximately linear parts. The temperature in the middle region is found to be higher than on the two outside regions as anticipated by comparison to the the previous results shown in Fig.~\ref{fig:QP1}. In both cases the temperature is larger when the pre-quench Hamiltonian over the region of interest is (almost) flat compared to the case where pre-quench Hamiltonian has (almost) linear profile.

We also see that the distributions for both middle and side regions in this case are not exactly thermal as reflected in the energy dependence of $T(E)$ shown in the insets of Fig.~\ref{fig:QP2}.
However, the distribution for the middle region is much more closer to an exact Fermi--Dirac distribution, with a fitting error of $\chi\sim 10^{-3}$ compared to the side regions, for which the error is $\chi\sim 10^{-2}$.
Similarly, for the more generic case of quenching from $P_1$ to $P_2$ we expect thermalization to  
be closer to ideal in the regions where the pre-quench profile is almost flat. 
In that case the temperature is mainly determined by the post-quench profile, and approximately scales as $T \propto P_2$ due to the increase in surface gravity by quenching to $P_2$.
Indeed, the variations in effective temperature observed following different types of quenches in Figs.~\ref{fig:QP1} and \ref{fig:QP2} can be understood in terms of changes in the effective surface gravity. In the gravitational analogy, this surface gravity corresponds to the spatial gradient of the hopping amplitude near the horizon. A steeper gradient in the hopping profile results in a larger surface gravity, and consequently, a higher effective temperature.

\subsection{Off-diagonal correlations}
In this section, we examine how the slight deviations from an exact thermal distribution are reflected in the off-diagonal correlations between different energies, denoted as $\langle b_{E}^\dagger b^{\phantom\dagger}_{E'} \rangle$.

As shown in Fig.~\ref{fig:correlations}(a), when the expectation values for the occupation numbers of post-quench modes are exactly thermal, there are no off-diagonal correlations between different energy modes. Conversely, slight deviations from the exact thermal distribution, such as those occurring when quenching from the sine models with $P=1$ to $P=2$, are accompanied by deviations from a purely diagonal form. This is illustrated in Fig. \ref{fig:correlations}(b), where small nonzero off-diagonal correlations appear near the zero modes, $E, E' \sim 0$.

As we move away from the zero modes to a region in which the diagonal terms corresponding to the expectation values of the occupation numbers closely approach zero or one, the off-diagonal terms are suppressed, as expected. Similar behavior, characterized by simultaneous deviations from the thermal form of the diagonal terms and small non-zero off-diagonal terms around zero energy, is observed in other cases, such as quenching from $P=1$ to $P=3$, and other such quenches.

\subsection{Connection to Harper's model}

It might seem intuitive that the exact mapping between the sine and Harper models would allow for the engineering of similar thermalization phenomena in a quantum Hall system by simply and suddenly changing the magnetic flux. However, unlike the sine model, where quench-induced thermalization can be directly analyzed, such a protocol does not straightforwardly lead to thermalization in the Harper-model description of the quantum Hall system.

While all other results presented in this work rely on the exact mapping between the sine model and Harper's equation for a single integer $P$, the situation becomes more complex when considering quench-induced thermalization associated with the creation of new horizons. This is because such scenarios necessarily involve two distinct Hamiltonians (in either the sine or Harper-model representations), corresponding to pre- and post-quench systems with different values of $P$. 

Given that the precise form of the mapping depends explicitly on $P$, applying the transformation associated with the post-quench Hamiltonian from the sine to the Harper model requires applying the same transformation to the pre-quench ground state, which corresponds to a different value of $P$. As a consequence, the resulting initial state in the Harper-model representation acquires a highly nontrivial and artificial structure, bearing little resemblance to the ground state of a simple Harper model.

\begin{figure}
    \centering
    \includegraphics[width=8.6cm]{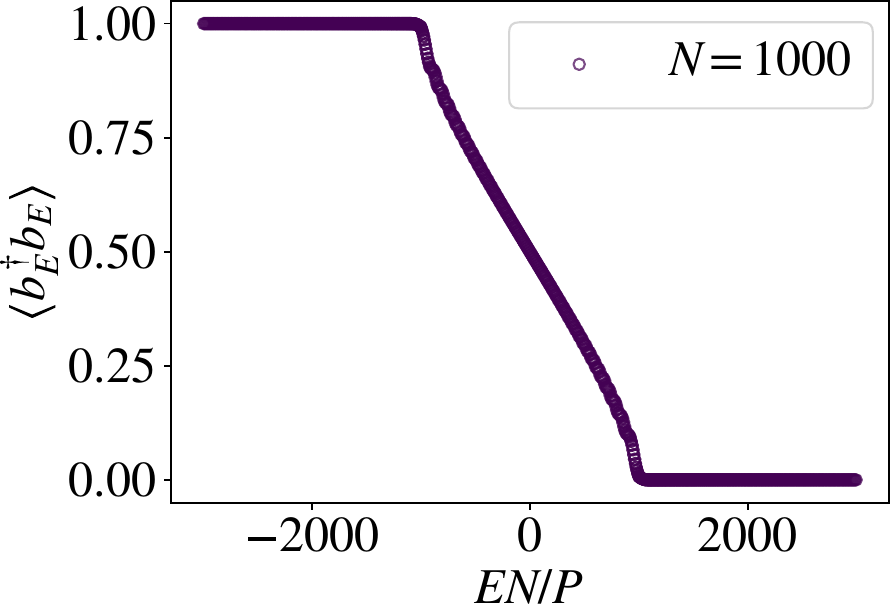}
    \caption{Expectation values of the occupation number $\langle b_E^\dagger b^{\phantom\dagger}_E\rangle$ for the post-quench eigenmodes following a direct quench from the uniform-hopping Hamiltonian to the Harper-model Hamiltonian. The resulting distribution is clearly far from a thermal Fermi--Dirac form.}
    \label{fig:nonthermal-harper}
\end{figure}

On the other hand, if the mapping from the sine model to Harper’s model is applied only to the post-quench Hamiltonian, thermalization does not occur. This situation corresponds, for example, to a quench from a uniform-hopping model, which is equivalent to Harper’s equation in the absence of a magnetic field, to Harper’s equation with a finite magnetic field. In this case, the resulting state is highly non-thermal, as illustrated in Fig.~\ref{fig:nonthermal-harper}.

For this reason, while the emergence of effective Unruh temperatures can be explicitly demonstrated and analyzed within the sine model, their realization in the quantum Hall setting cannot be addressed on the same footing. Observing any quench-induced Unruh-like thermal behavior in the Harper or quantum Hall framework would therefore require careful and nontrivial engineering of the initial state, going beyond the simple flux-quench scenario.

\section{Discussion and Summary}
We used a direct connection between quantum Hall states in a lattice described by Harper's model and a 1D sinusoidal hopping model to explore analog gravitational features in both. We first showed the correspondence between the two models by studying semiclassical equations of motions and wave-packet dynamics on a lattice. It has been found that both Harper's model and the sine model host synthetic horizons, where quasiparticles slow down exponentially.

Our findings regarding the quantum Hall system suggest a direct generalization: synthetic horizons and thermalization may be expected to arise in {\it any} quantum lattice system where a saddle point in the semiclassical continuum phase portrait causes classical particles to slow down exponentially. Similar to the findings presented here, the dynamics of coherent wave packets under the full quantum evolution may then be expected to follow the semiclassical path, until a point where the finite lattice spacing forces the wave packets to disintegrate.

Furthermore, we demonstrated the potential of \emph{the sine model} to act as a fundamental tool for studying Unruh-like effects and thermalization resulting from sudden changes in the periodicity of the sinusoidal hopping profile. That the post-quench mode occupation is precisely thermal is not a universal occurrence; it manifests only when the so-called entanglement Hamiltonian of the pre-quench system closely approximates the actual Hamiltonian of the post-quench system~\cite{Mertens2022}. 
The entanglement Hamiltonian in the sine model {\it equals} the post-quench Hamiltonian for a quench from uniform to a single sine hopping profile ($P=0$ to $P=1$), and we confirmed the resulting purely thermal occupation profile. 

We emphasize that the notion of \emph{thermalization} discussed here differs from the more generic term in the context of many-body systems, where thermalization arises from interactions and slow relaxation toward equilibrium. In our setting, thermal-like behavior emerges immediately after a carefully designed quench protocol, rooted in a gravitational analogy to the Unruh effect. This mechanism relies on the structure of the post-quench Hamiltonian, which mimics the entanglement Hamiltonian of the pre-quench system and features synthetic horizons. Although distinct from interaction-driven thermalization, the resulting mode occupations follow a thermal distribution, justifying the use of the term ``thermalization” in this context. This perspective opens avenues for probing emergent temperatures through mode statistics or energy exchange with coupled systems.
\par
While this effective thermal description is most transparent immediately after the quench, the subsequent time evolution is inherently non-equilibrium. After the quench, the system undergoes transient dynamics before reaching its long-time behavior. Since the model is noninteracting and integrable, the system may not relax to a conventional thermal state at long times. Nevertheless, in the presence of even very weak interactions which break integrability, genuine thermalization can be eventually reached. In that case, simple thermodynamic considerations imply that the temperature of the final equilibrium state must coincide with the effective temperature characterizing the immediate post-quench distribution, since the system is closed and energy is conserved. Although a detailed numerical analysis of this long-time relaxation is beyond the scope of the present work, this expectation provides a consistent interpretation of the effective thermal behavior observed after the quench.
\par
From an experimental standpoint, both the sine and Harper models can be readily realized and controlled in cold atom setups \cite{Aidelsburger2013,Aidelsburger2015}. The formation of horizons and the exponential slowdown of wave packets can also be observed in these systems by dynamically generating wave packet-like excitations and allowing them to evolve. Furthermore, the Unruh effect predicted for the sine model upon performing sudden quenches, as a fully quantum phenomenon that depends on the multi-particle fermionic nature of the model, can also be explored in cold atom experiments. Moreover, since the wave packet dynamics is fundamentally a single-particle phenomenon, it does not depend on the fermionic statistics of the underlying system and can therefore also be studied in photonic metamaterials or electrical circuits that effectively model the sinusoidal hopping of the sine model or the potential modulation in the case of the Harper model.
\par
The sine model and its connection with Harper's equation thus provide a convenient platform for exploring horizon physics and thermalization in analog gravity, in which particle dynamics as well as the causes, consequences, and non-universal features of thermalization can be studied.

\subsection*{Acknowledgments} This work was supported by the Deutsche Forschungsgemeinschaft (DFG, German Research Foundation) through the Sonderforschungsbereich SFB 1143, grant No. YE 232/2-1, and under Germany's Excellence Strategy through the W\"{u}rzburg-Dresden Cluster of Excellence on Complexity and Topology in Quantum Matter -- \emph{ct.qmat} (EXC 2147, project-ids 390858490 and 392019). A.G.M. acknowledge Jane and Aatos Erkko Foundation for financial support.

\appendix

\section{Continuum models}
\label{sec1}

\subsection{Harper's model}

We start by considering Harper's equation~\cite{harper1955single}: 
\begin{align}
    \psi_{n-1} + 2 \cos(2 \pi \alpha n + \theta) \psi_n + \psi_{n+1} = \epsilon \psi_n.
    \label{eq:Harper-SM}
\end{align}
where the state $\psi_n$ is defined on discrete lattice sites $n \in [0,N-1]$, the energy is denoted by $\epsilon$, and $\theta \in \mathbb{R}$ is the quasi-momentum corresponding to the Fourier transformed spatial direction. Finally, $\alpha=P/N$ is equal to the magnetic flux through a lattice cell divided by the magnetic flux quantum. 

By going to very large systems $N\to\infty$ and $\alpha\to 0$ we can construct the semiclassical limit of the Harper's equation.
The $\psi_n$ eigenstate is replaced by $\psi(y)$ with $\psi(n)=\psi_n$.
For large enough $N$ we can think of $\psi(y)$ as a continuous differentiable function.

The Harper's equation in this approximation becomes
\begin{align}
\psi(y+1)+\psi(y-1)+2\cos(2\pi\alpha y + \theta)\psi(y) = \epsilon \psi(y).
\end{align}

We introduce the position operator $\hat{y}$ and the conjugate momentum $\hat{q}$ with commutation relation $[\hat{y},\hat{q}]=i$. Translations are generated by $\hat{q}$ and the shifted wave function can be expressed as
\begin{equation}
    \psi(y+1) = \mathrm{e}^{i\hat{q}}\psi(y).
\end{equation}

With this, the Harper's equation can be expressed as
\begin{align}
    \hat{\cal{H}}_H \psi(y) & = \epsilon \psi(y),
\end{align}
with the corresponding Hamiltonian
\begin{align}
    \hat{\cal{H}}_H  & = 2\cos(\hat{q}) + 2\cos(2\pi \alpha \hat{y} + \theta). 
    \end{align}

For this Hamiltonian, we can calculate the semiclassical dynamics. 
The Heisenberg equations of motion yield 
\begin{align*}
    \dv{\hat{y}}{t} & = i[\hat{\cal{H}}_H ,\hat{y}] = - 2 \sin(\hat{q}), \\
    \dv{\hat{q}}{t} & = i[\hat{\cal{H}}_H ,\hat{q}] = 4 \pi \alpha \sin(2\pi\alpha \hat{y} + \theta).
\end{align*}
Using these equations and the Ehrenfest theorem we can get the semiclassical dynamics of $y=\langle\hat{y}\rangle$ and $q=\langle\hat{q}\rangle$ by replacing the operators with their expectation value in the Heisenberg equations of motion. We can solve these equations using
\begin{equation}
\dv{q}{y} = -2\pi\alpha \frac{\sin(2\pi\alpha y+\theta)}{{\sin(q)}},
\end{equation}
which has linear solutions (this will lead to solutions on the separatrices)
\begin{align}
    q = 2\pi\alpha y + \theta  + (2m+1)\pi, 
\end{align}
where $m\in\mathbb{Z}$.
With this, the time evolution of the position $y(t)$ becomes 
\begin{align}
    y(t) = \frac{1}{\pi \alpha} \arctan\Big(e^{-4\pi\alpha(t-t_0)}\Big) - \frac{\theta}{2\pi\alpha} - \frac{2m+1}{2\alpha}.
\end{align}
This solution shows exponential slowing down for $t \to \infty$, at $y = -\theta/(2 \pi \alpha) - \frac{2m+1}{2\alpha}$. Similarly, the time dependence for $q$ becomes 
\begin{align*}
    q(t) = 2\arctan\Big(e^{-4\pi \alpha (t-t_0)} \Big) + 2n\pi,
\end{align*}
where $n\in\mathbb{Z}$.
At $t \to \infty$ this reduces to $2n\pi$. 
Away from the separatrix, different solutions exist, as shown in Fig. 1 in the main text.

For example, at $q = q'$ and $y= -\theta/(2\pi\alpha) + y'$ for small $q', y'$ the Hamiltonian becomes 
\begin{align*}
   \hat{\cal{H}}_H  \approx 4+(q')^2 + (2\pi\alpha y')^2.
\end{align*}
This is the formula for a quantum harmonic oscillator which give the circular orbits in Fig.~1 of the main text.

\subsection{Sine model}
The sine model is defined as
\begin{align}
    2\sin(\pi n \alpha )\phi_{n-1} + 2\sin(\pi(n + 1)\alpha )\phi_{n+1} = \epsilon \phi_n,
\end{align}

In the semiclassical limit, similarly to the previous section (now position and the conjugate momentum are $\hat{x}$ and $\hat{p}$) for the Hamiltonian we get \footnote{It should be noted that this Hamiltonian is manifestly Hermitian as the second term $\sin(\pi (\hat{x}+1)\alpha )e^{i\hat{p}}$
can be also re-written as $e^{i\hat{p}} \sin(\pi \hat{x}\alpha)$ which is the Hermitian conjugate of the first term.
}

\begin{align}
\label{eq:Hsfull}
    \hat{\cal{H}}_S = 2\sin(\pi \hat{x}\alpha)e^{-i\hat{p}} + 2\sin(\pi (\hat{x}+1)\alpha)e^{i\hat{p}}. 
\end{align}

For large enough systems where $\alpha\to 0$ we can approximate the Hamiltonian as
\begin{align}\label{eq:H_approx}
    \hat{\cal{H}}_S \approx 4\sin(\pi \hat{x}\alpha)\cos{(\hat{p})}.
\end{align}
We note that, in contrast to the exact expression given in Eq. \eqref{eq:Hsfull}, the approximate form \eqref{eq:H_approx} is non-Hermitian. This form arises by approximating $\alpha(\hat{x} + 1) \approx \alpha \hat{x}$, which effectively neglects the non-commutativity between $\hat{x}$ and $\hat{p}$. Such an approximation is justified in the semiclassical limit, where operator ordering becomes insignificant and the lack of Hermiticity has no impact on the physical results.

Using the Heisenberg equations of motion for the sine model we get
\begin{align}
    \dv{\hat{x}}{t} & = i[\hat{\cal{H}}_S ,\hat{x}] = -4 \sin(\pi \hat{x}\alpha)  \sin(\hat{p}), \\
    \dv{\hat{p}}{t} & = i[\hat{\cal{H}}_S ,\hat{p}] = -4\alpha \pi \cos(\pi \hat{x}\alpha )\cos(\hat{p}).
\end{align}

This coincides with the result known for a model with linearly increasing hopping strength~\cite{Mertens2022}, in which a wave packet centered at $p\equiv\langle\hat{p}\rangle = \pm\pi/2 $ has constant momentum and a trajectory given by
\begin{align}
    x(t) = \frac{2}{\pi \alpha} \arctan(e^{\mp 4 \pi \alpha (t-t_0) } ) + \frac{2m}{\alpha},
\end{align}
where $m\in\mathbb{Z}$.
For $t \to \infty$ there are multiple horizons at $x_h = x(t= \infty) = m'/\alpha$ for $m'\in\mathbb{Z}$.

We can write an effective continuum model for momenta close to $p = \pm \pi/2$ and near the horizons.
Taking $\alpha = 1/N$ we have the horizons at $x=0$ and $x=N$. 
Close to the left and right horizon we introduce small parameters $x'$ and $p'$
\begin{align}
    & \hat{x} = \hat{x}' &  \hat{p} &= \pi/2 + \hat{p}', \\
    & \hat{x} = N+\hat{x}'' &  \hat{p} &= - \pi/2 + \hat{p}''.
\end{align}
Substituting to Eq.~\eqref{eq:Hsfull} the effective Hamiltonian becomes
\begin{align}
    \hat{\cal{H}}_S' &\approx
    - 2\pi\alpha (\hat{x}'\hat{p}' + \hat{p}'\hat{x}'),\\
        \hat{\cal{H}}_S'' &\approx
     2\pi\alpha (\hat{x}''\hat{p}'' + \hat{p}''\hat{x}'').
\end{align}
This is proportional to the Keating-Berry operator.

\section{Transformation from Harper's equation to the sine model} \label{sec2}

Following the transformations introduced in Ref.~\cite{krasovsky1999bethe}, we will show how the Harper's model can be transformed to the sine model. 
We can re-write Harper's Eq. \eqref{eq:Harper} as
\begin{align} \label{eq:mu}
    \psi_{n-1} + \left(q^{2n} e^{i\theta} + q^{-2n}e^{-i\theta} \right)\psi_n + \psi_{n+1} = \epsilon \psi_n ,
\end{align}
where $q \equiv e^{i\pi\alpha}$. 
We consider periodic boundary conditions given by $\psi_N =\psi_0$.
We define the following transformation
\begin{align}
\label{eq:T1}
    \psi_n = \frac{1}{\sqrt{N}}\sum_{k=0}^{N-1} q^{2nk}  e^{ik(\theta - \theta^\prime)} \chi_k ,
\end{align}
with inverse transformation
\begin{align}
\label{eq:T1I}
    \chi_k = \frac{1}{\sqrt{N}} \sum_{n=0}^{N-1} q^{-2nk}e^{-ik(\theta - \theta^\prime)}  \psi_n,
\end{align}
where $\theta^\prime\in\mathbb{R}$ is introduced as a free parameter (its function will become evident later).
We note that the inverse transformation \eqref{eq:T1I} implies a periodicity condition
\begin{align}\label{eq:periodicity}
    \chi_{k+N} &= e^{-i N(\theta -\theta^\prime)}\chi_k,  
 \end{align}
and in particular $\chi_N = e^{-i N(\theta -\theta^\prime)}\chi_0$.
The inverse transformation is consistent only when $N$ and $P$ are co-primes in $\alpha=P/N$. The reason is that we require
the so-called completeness condition
\begin{align}
    \sum_{k=0}^{N-1} q^{2nk} = N\delta_{n,0},
\end{align}
which in turn is valid only when there is no other integer than $n=0$ in the range of $[0,N-1]$ for which $q^{2n}=1$. If $P$ and $N$ have a common divisor larger than 1 then, there will be other $n$'s for which $q^{2n}=1$ and the completeness is not satisfied.

Inserting Eq. \eqref{eq:T1} into Eq. \eqref{eq:mu} yields
\begin{align*}
    \sum_{k=0}^{N-1} & \bigg[ q^{2(n-1)k} e^{ik(\theta -\theta^\prime)} +  q^{2n(k+1)}e^{ik(\theta-\theta^\prime) +i \theta}  \\
     + & q^{2n(k-1)}e^{ik(\theta -\theta^\prime) - i\theta}
     + q^{2(n+1)k} e^{ik(\theta -\theta^\prime)} \bigg] \chi_k
     \\
     & = \epsilon \sum_{k=0}^{N-1} q^{2nk}  e^{ik( \theta  -\theta^\prime)} \chi_k.
\end{align*}
The middle two terms can be rewritten by defining $k'= k \pm 1$. This gives a sum that is shifted by $\pm 1$ from the desired domain. 
This leaves us with the equation 
\begin{align}
    & e^{i\theta^\prime}\chi_{k-1} + (q^{2k} + q^{-2k})\chi_k + e^{-i \theta^\prime}\chi_{k+1} = \epsilon \: \chi_k.
    \label{eq:phi}
\end{align}

We apply a second transformation, given by
\begin{align}
\label{eq:T2}
    \chi_k = \frac{1}{\sqrt{N}}\sum_{n=0}^{N-1} q^{(k+1/2)^2} q^{2kn} \zeta_n.
\end{align}
with inverse transformation
\begin{align}
\label{eq:T2I}
    \zeta_n = \frac{1}{\sqrt{N}}\sum_{k=0}^{N-1} q^{-(k+1/2)^2}q^{-2kn} \chi_k. 
\end{align}
From Eq. \eqref{eq:T2} we can easily see that
\begin{align}
    \chi_N = q^{(N+1/2)^2-(1/2)^2}\chi_0,
\end{align}
which in combination with the the periodicity condition, Eq. \eqref{eq:periodicity}, it leads to
\begin{align}
    e^{-iN(\theta-\theta^\prime)} = q^{(N+1/2)^2-(1/2)^2},
\end{align}
and thus
\begin{align}
\label{eq:Ntheta}
    N(\theta - \theta^\prime) + \pi P (N+1) = 0  \text{ mod } 2\pi.
\end{align}
Also, the inverse transformation is consistent as long as $P$ and $N$ are co-primes.
Hence, for even and odd $N$, we have
\begin{align}
\label{eq:even-theta}
\theta^\prime_e &= \theta+\pi\alpha+\frac{2\pi m}{N},\\
\label{eq:odd-theta}
\theta^\prime_o &= \theta+\frac{2\pi m}{N},
\end{align}
for an arbitrary integer $m\in\mathbb{Z}$.
Inserting Eq. \eqref{eq:T2} into Eq. \eqref{eq:phi} gives
\begin{align}
\nonumber
    \sum_{n=0}^{N-1} & \bigg[ q^{(k-1/2)^2}q^{2(k-1)n}e^{i\theta^\prime} + q^{(k+1/2)^2}q^{2k(n+1)} 
    \\ \nonumber +&  q^{(k+1/2)^2}q^{2k(n-1)} + q^{(k+3/2)^2}q^{2(k+1)n} e^{-i\theta^\prime}\bigg]\zeta_n 
    \\ &= \epsilon\sum_{n=0}^{N-1} q^{(k+1/2)^2}q^{2kn}\zeta_n.
\end{align}
For the first term we take $n' = n - 1$, which yields
\begin{align}
\nonumber
    & \sum_{n'=-1}^{N-2} q^{k^2 - k + 1/4} q^{2kn'-2n'-2 + 2k} e^{i\theta^\prime} \zeta_{n'+1}=
    \\ \nonumber = & \sum_{n'=-1}^{N-2} q^{(k+1/2)^2}q^{2kn'}q^{-2(n'+1)}e^{i\theta^\prime}\zeta_{n'+1}=
        \\ = & \sum_{n'=0}^{N-1} q^{(k+1/2)^2}q^{2kn'}q^{-2(n'+1)}e^{i\theta^\prime}\zeta_{n'+1}.
\end{align}
The last step works by taking the choice $\zeta_0 = \zeta_N$. The other three terms can be similarly shifted to yield 
\begin{align}
    & \left(1 + e^{-i\theta^\prime}q^{2n} \right) \zeta_{n-1}
     +\left(1 + e^{i\theta^\prime}q^{-2(n+1)} \right)\zeta_{n+1} = \epsilon \zeta_n. 
     \label{eq:zeta}
\end{align}

Now,  we further apply another transformation
\begin{align}
\label{eq:zetatrans}
    \zeta_{n} = q^{n(n+1)/2} e^{-i n\theta^\prime/2} \phi_n ,
\end{align}
which after some algebra results in the following equation:
\begin{align}
    \cos\left( \pi \alpha n
    -\frac{\theta^\prime}{2}
    \right) \phi_{n-1} + \cos\left(\pi \alpha (n + 1) - \frac{\theta^\prime}{2}\right) \phi_{n+1} = \frac{\epsilon}{2} \phi_n ,
\end{align}
or equivalently
\begin{align}
    &\sin\left( \pi \alpha n
    +\frac{\pi-\theta^\prime}{2}
    \right) \phi_{n-1} \nonumber\\
    &\qquad + \sin\left(\pi \alpha (n + 1) +\frac{\pi-\theta^\prime}{2}\right) \phi_{n+1} = \frac{\epsilon}{2} \phi_n ,
\end{align}
which is the desired sine model.

The boundary conditions from Eq.~\eqref{eq:zetatrans} are $\phi_{N} = \phi_0 q^{-N(N+1)/2} e^{i N\theta^\prime/2}$ and $\phi_{-1} = \phi_{N-1} q^{N(N-1)/2} e^{-i N\theta^\prime/2}$.

The transformation is particularly simple for $\theta^\prime=\pi$ in this case $\phi_{-1}$ and $\phi_N$ are disconnected from the rest of the chain, meaning that the boundary condition can be chosen arbitrarily.
Therefore, using Eqs. \eqref{eq:even-theta} and \eqref{eq:odd-theta}, the above condition is satisfied by
\begin{align}
    \theta &= \frac{\pi}{N}(N-P-2m) & &\text{for even}\ N,\\
    \theta &= \frac{\pi}{N}(N-2m) & &\text{for odd}\ N.
\end{align}
Since $P$ and $N$ cannot be both even (they are co-primes), both results above simplify to
\begin{align}
    \theta = (2m'+1)\frac{\pi}{N},
\end{align}
with $m'\in\mathbb{Z}$.

To summarize, the requirements for the transformations used in this case to work, are:
\begin{enumerate}
    \item $P$ and $N$ are co-primes.
    \item  $\theta = (2m'+1)\frac{\pi}{N}$ with $m'\in\mathbb{Z}$
\end{enumerate}
These lead to the sine model used in the main paper
\begin{align}
    &2\sin\left( \pi \alpha n
    \right) \phi_{n-1} + 2\sin\left(\pi \alpha (n + 1)\right) \phi_{n+1} = {\epsilon} \phi_n.
\end{align}

\section{Calculating the distribution of post-quench modes \label{sec3}}

In this section, we present the details of obtaining the distribution (expectation values of occupancy) of the post-quench modes. Let us denote the set of all single-particle eigenstates of the pre-quench Hamiltonian as $\{ \ket{\phi_\omega^{\rm pre}} \}$, which spans the full Hilbert space of size $N$, equal to the number of lattice points. The post-quench Hamiltonian, however, results in multiple causally-disconnected regions, with no Hamiltonian coupling terms between different regions. 

After the quench, the full Hilbert space of $N$ lattice points can be considered decomposed as $\mathcal{H}_0 = \bigoplus_{j=1}^{m} \mathcal{H}_j$, where each subspace corresponds to a different post-quench sub-region $j$. For each sub-region, we have separate sets of eigenstates $\{ \ket{\psi_E^{{\rm post},j}} \}$. Note that we assume the indices $\omega$ and $E$ uniquely identify different eigenstates in both cases. In the context of a 1D sine model, we do not expect extensive degeneracies. However, in the event of degeneracies, we implicitly assume that $\omega$ and $E$ are accompanied by an additional quantum number, allowing us to label all eigenstates accordingly.

Since both $\{ \ket{\phi_\omega^{\rm pre}} \}$ and $\bigoplus_j \{ \ket{\psi_E^{{\rm post},j}} \}$ span the full Hilbert space $\mathcal{H}_0$, there exists a unitary transformation $U$ that connects these sets of eigenstates:

\begin{align}
    &\ket{\psi_E^{{\rm post},j}} = \sum_{\omega} U_{(E,j);\omega} \ket{\phi_\omega^{\rm pre}}, \label{eq:U}\\
    &\ket{\phi_\omega^{\rm pre}} = \sum_{E,j} U^\ast_{\omega;(E,j)} \ket{\psi_E^{{\rm post},j}}.
\end{align}
As mentioned earlier, the index $\omega$ and the double index $(E,j)$ each take $N$ different values.

Next, we demonstrate how to calculate the expectation values $\langle b^\dag_{E,j} b^{\phantom\dagger}_{E',j} \rangle \equiv \bra{GS} b^\dag_{E,j} b^{\phantom\dagger}_{E',j} \ket{GS}$ for the post-quench modes. These operators, $b^\dag_{E,j}$ and $b^{\phantom\dagger}_{E',j}$, correspond to different eigenstates in subregion $j$, with respect to the ground state of the pre-quench Hamiltonian:
\begin{align}
    \ket{GS} = \prod_{\omega < 0} c_{\omega}^\dag \ket{0}.
\end{align}

Here, $c_{\omega}^\dag$ is the creation operator for a pre-quench energy eigenstate. Using Eq. \eqref{eq:U}, we can immediately express the transformation between post- and pre-quench operators as:
\begin{align}
    &b_{E,j}^\dag = \sum_{\omega} U_{(E,j);\omega} \: c_{\omega}^\dag, \nonumber \\
    &b_{E,j}^{\phantom\dagger} = \sum_{\omega} U^\ast_{(E,j);\omega} \: c^{\phantom\dagger}_{\omega}.
\end{align}
Substituting these expressions into the expectation values of the post-quench mode correlations, we obtain:
\begin{align}
\langle b^\dag_{E,j} b^{\phantom\dagger}_{E',j} \rangle &= \sum_{\omega,\omega'} U_{(E,j);\omega} U^\ast_{(E',j);\omega'} \bra{GS} c_{\omega}^\dag c_{\omega'}^{\phantom\dagger} \ket{GS} \nonumber \\
&= \sum_{\omega<0} U_{(E,j);\omega} U^\ast_{(E',j);\omega}
\end{align}
This forms the basis for our numerical evaluation of the correlations and expectation values of mode occupations presented in the paper.

\begin{figure}[!]
    \centering
\includegraphics[width=0.95\linewidth]{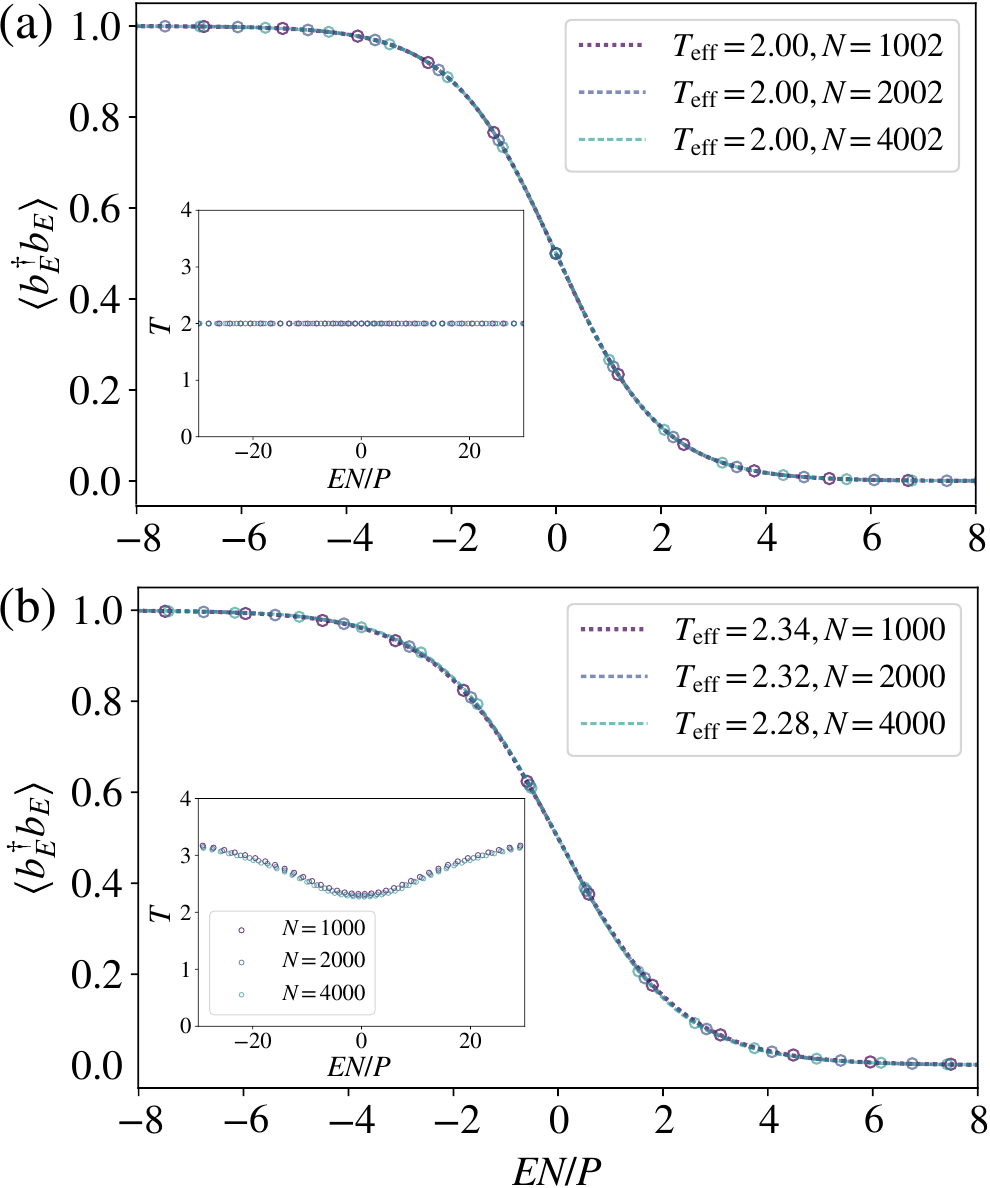}
    \caption{Illustration of the even-odd dependence on the size of post-quench sub-regions. Panels (a) and (b) show the average occupation number and their thermal fit when we quench from the uniform hopping model ($P=0$) to the sine model ($P=1$), but with sizes such that the corresponding subregion size ($N/2$ in this case) is odd for panel (a) and even for panel (b).
    \label{fig:even-odd}}
\end{figure}

\subsection{Even-odd effect dependence on sub-region size}
We finally compare the results for a quench from a uniform hopping model ($P=0$) to a sine hopping model ($P=1$), when considering either even or odd numbers of lattice points inside each post-quench sub-region.
Because the initial region in this quench scenario is twice as large as the initial one, the even and odd cases correspond to the initial lattice sizes being multiples of four and two (but not four), respectively. The main difference between these two cases is that with an odd number of sites and particle-hole symmetry, there exists a zero-energy mode, whereas in the even case, there is no zero-energy mode and instead we have a pair of states with small energies $\pm E_1$. Consequently, in the even case, the exact thermal form is not achieved for any finite lengths of the chain, as shown in Fig. \ref{fig:even-odd}.

It can be inferred from the trend of fitted effective temperatures for different $N$ in Fig. \ref{fig:even-odd}(b), that increasing $N$ leads to a slight decrease in the variation of effective temperatures. Hence, for very large $N$ we expect the even-odd effect to disappear and the results for the two types of lattice sizes to reconcile. This type of even-odd effect, where the results slightly differ between cases where the size of the post-quench subregion of interest is even or odd, is a generic behavior of these models. We also found it to be present when we quench between two different values of the magnetic flux corresponding to two different values of $P$.

\bibliography{ref}

\begin{thebibliography}{65}
\bibitem [{foo()}]{footnote-2}%
  \BibitemOpen
  \href@noop {} {}\bibinfo {note} {We use absolute fitting error $\chi^2 =
  \sum_i [f_i - f_{\rm FD}(E_i,T_{\rm eff})]^2 $ with $f_i$ denoting the
  numerically calculated post-quench occupation and $f_{\rm FD}$ the
  Fermi-Dirac function with a fitted temperature $T_{\rm eff}$.}\BibitemShut
  {Stop}%
\end{thebibliography}

\end{document}